\documentclass{article}

\PassOptionsToPackage{numbers, compress}{natbib}
\usepackage[final]{neurips_2020}

\usepackage[utf8]{inputenc} 
\usepackage[T1]{fontenc}    
\usepackage{hyperref}       
\usepackage{url}            
\usepackage{booktabs}       
\usepackage{amsfonts}       
\usepackage{nicefrac}       
\usepackage{microtype}      
\usepackage[pdftex]{graphicx}
\usepackage{xcolor}
\usepackage{amsmath,bm}
\usepackage{mathtools}
\usepackage{tabularx}

\usepackage{algorithm} 
\usepackage{algpseudocode}

\DeclarePairedDelimiterX{\infdivx}[2]{\big[}{\big]}{%
  #1\;\delimsize\|\;#2%
}
\newcommand{\KL}{D_{\text{KL}}\infdivx}
\newcommand{\E}[1]{\mathop{\mathbb{E}_{#1}}}
\usepackage[colorinlistoftodos]{todonotes}

\newcommand\subref[2]{\hyperref[#1]{\ref*{#1}#2}}

\title{Deep active inference agents using Monte-Carlo methods}

\author{%
  Zafeirios Fountas\thanks{Corresponding author} \\
  Emotech Labs \& \\
  WCHN, University College London \\
  \texttt{f@emotech.co} \\
  \And
  Noor Sajid \\
  WCHN, University College London \\
  \texttt{noor.sajid.18@ucl.ac.uk} \\
  \And
  Pedro A.M. Mediano \\
  University of Cambridge \\
  \texttt{pam83@cam.ac.uk} \\
  \And
  Karl Friston \\
  WCHN, University College London \\
  \texttt{k.friston@ucl.ac.uk} \\
}

\begin{document}
\maketitle

\begin{abstract}
Active inference is a Bayesian framework for understanding biological intelligence. The underlying theory brings together perception and action under one single imperative: minimizing free energy. However, despite its theoretical utility in explaining intelligence, computational implementations have been restricted to low-dimensional and idealized situations. In this paper, we present a neural architecture for building deep active inference agents operating in complex, continuous state-spaces using multiple forms of Monte-Carlo (MC) sampling. For this, we introduce a number of techniques, novel to active inference. These include: $i)$ selecting free-energy-optimal policies via MC tree search, $ii)$ approximating this optimal policy distribution via a feed-forward `habitual' network, $iii)$ predicting future parameter belief updates using MC dropouts and, finally, $iv)$ optimizing state transition precision (a high-end form of attention). Our approach enables agents to learn environmental dynamics efficiently, while maintaining task performance, in relation to reward-based counterparts. We illustrate this in a new toy environment, based on the dSprites data-set, and demonstrate that active inference agents automatically create disentangled representations that are apt for modeling state transitions. In a more complex Animal-AI environment, our agents (using the same neural architecture) are able to simulate future state transitions and actions (i.e., plan), to evince reward-directed navigation - despite temporary suspension of visual input. These results show that deep active inference -- equipped with MC methods -- provides a flexible framework to develop biologically-inspired intelligent agents, with applications in both machine learning and cognitive science.
\end{abstract}

\section{Introduction}
A common goal in cognitive science and artificial intelligence is to emulate biological intelligence, to gain new insights into the brain and build more capable machines. A widely-studied neuroscience proposition for this is the free-energy principle, which views the brain as a device performing variational (Bayesian) inference \cite{friston2010free,friston2019free}. Specifically, this principle provides a framework for understanding biological intelligence, termed active inference, by bringing together perception and action under a single objective: minimizing free energy across time \cite{learningpaper,friston2017active,friston2017graphical,pezzulo2018hierarchical,da2020active}. However, despite the potential of active inference for modeling intelligent behavior, computational implementations have been largely restricted to low-dimensional, discrete state-space tasks \cite{Parr:2017:Uncertainty,deeptemporal,sajid2019demystifying,hesp2019deeply}. 

Recent advances have seen deep active inference agents solve more complex, continuous state-space tasks, including Doom \cite{Cullen}, the mountain car problem \cite{active_rl,ueltzhoffer2018deep,ccatal2019bayesian}, and several tasks based on the MuJoCo environment \cite{tschantz2019scaling}, many of which use amortization to scale-up active inference \cite{active_rl,ueltzhoffer2018deep,ccatal2019bayesian,millidge2020deep}. A common limitation of these applications is a deviation from vanilla active inference in their ability to plan. For instance, Millidge~\cite{millidge2020deep} introduced an approximation of the agent's \textit{expected free energy} (EFE), the quantity that drives action selection, based on bootstrap samples, while Tschantz \emph{et al.}~\cite{tschantz2019scaling} employed a reduced version of EFE. Additionally, since all current approaches tackle low-dimensional problems, it is unclear how they would scale up to more complex domains. Here, we propose an extension of previous formulations that is closely aligned with active inference \cite{friston2017active,deeptemporal} by estimating all EFE summands using a single deep neural architecture.

Our implementation of deep active inference focuses on ensuring both scalability and biological plausibility. We accomplish this by introducing Monte-Carlo (MC) sampling -- at several levels -- into active inference. For planning, we propose the use of MC tree search (MCTS) for selecting a free-energy-optimal policy. This is consistent with planning strategies employed by biological agents and provides an efficient way to select actions (see Sec.~\ref{sec:conclusion}). Next, we approximate the optimal policy distribution using a feed-forward ‘habitual’ network. This is inspired by biological habit formation, when acting in familiar environments that relieves the computational burden of planning in commonly-encountered situations. Additionally, for both biological consistency and reducing computational burden, we predict model parameter belief updates using MC-dropouts, a problem previously tackled with networks ensembles \cite{tschantz2019scaling}. Lastly, inspired by neuromodulatory mechanisms in biological agents, we introduce a top-down mechanism that modulates precision over state transitions, which enhances learning of latent representations. 

In what follows, we briefly review active inference. This is followed by a description of our deep active inference agent. We then evaluate the performance of this agent. Finally, we discuss the potential implications of this work.

\section{Active Inference}\label{activeinference}
Agents defined under active inference: $A)$ sample their environment and calibrate their internal generative model to best explain sensory observations (i.e., reduce surprise) and $B)$ perform actions under the objective of reducing their uncertainty about the environment. A more formal definition requires a set of random variables: $s_{1:t}$ to represent the sequence of hidden states of the world till time $t$, $o_{1:t}$ as the corresponding observations, $\pi=\{a_1,a_2, ...,a_T\}$ as a sequence of actions (typically referred to as `policy' in the active inference literature) up to a given time horizon $T \in \mathbb{N}^+$, and $P_{ \theta}(o_{1:t}, s_{1:t},a_{1:t-1})$ as the agent's generative model parameterized by $\theta$ till time $t$. From this, the agent's surprise at time $t$ can be defined as the negative log-likelihood  $-\log {P_{\theta}(o_t)}$. Through slight abuse of notation, $P_{\theta}(.)$ denotes distribution parameterisation by $\theta$ and $P(\theta)$ denotes use of that particular distribution as a random variable. See supplementary materials for definitions (Table \ref{gloss}).

To address objective $A)$ under this formulation, the surprise of current observations can be indirectly minimized by optimizing the parameters, $\theta$, using as a loss function the tractable expression: 
\begin{equation} \label{eq:F}
- \log P_{\theta}(o_{t}) \leq
 \mathop{\mathbb{E}_{Q_{\phi}(s_{t},a_{t})}}\big[ \log Q_{\phi}(s_{t},a_{t}) - \log P_{\theta}(o_{t}, s_{t},a_{t}) \big] ~ ,
\end{equation}
where $Q_{\phi}(s_t,a_t)$ is an arbitrary distribution of $s_t$ and $a_t$ parameterized by $\phi$. The RHS expression of this inequality is the \textit{variational free energy} at time $t$. This quantity is commonly referred to as negative \textit{evidence lower bound}~\cite{blei2017variational} in variational inference. Furthermore, to realize objective $B)$, the expected surprise of future observations $-\log {P(o_{\tau}|\theta,\pi)}$ where $\tau \geq t$ $-$ can be minimized by selecting the policy that is associated with the lowest EFE, $G$
\cite{parr2019generalised}: 
\begin{equation}
 \label{eq:G2}
G(\pi,\tau) = \mathop{\mathbb{E}_{P(o_{\tau}|s_{\tau},\theta)}\mathbb{E}_{Q_{\phi}(s_{\tau},\theta|\pi)}}\big[ \log Q_{\phi}(s_{\tau},\theta|\pi) - \log P(o_{\tau},s_{\tau},\theta|\pi) \big] ~ ,
\end{equation}
Finally, the process of action selection in active inference is realized as sampling from the distribution
\begin{equation} \label{eq:G1}
P(\pi) = \sigma\big(- G(\pi)\big) = \sigma\Big(- \sum_{\tau>t}{G(\pi,\tau)}\Big) ~ ,
\end{equation}
where $\sigma(\cdot)$ is the softmax function.

\section{Deep Active Inference Architecture}

In this section, we introduce a deep active inference model using neural networks, based on amortization and MC sampling.

Throughout this section, we denote the parameters of the generative and recognition densities with $\theta$ and $\phi$, respectively. The parameters are partitioned as follows: $\theta = \{\theta_o, \theta_s\}$, where $\theta_o$ parameterizes the observation function $P_{\theta_o}(o_t | s_t)$, and $\theta_s$ parameterizes the transition function $P_{\theta_s}(s_{\tau} | s_t, a_t)$. For the recognition density, $\phi = \{\phi_s, \phi_a\}$, where $\phi_s$ is the amortization parameters of the approximate posterior $Q_{\phi_s}(s_t)$ (i.e.,  the state encoder), and $\phi_a$ the amortization parameters of the approximate posterior $Q_{\phi_a}(a_t)$ (i.e., our habitual network).

\begin{figure}[t]
\includegraphics[width=\linewidth]{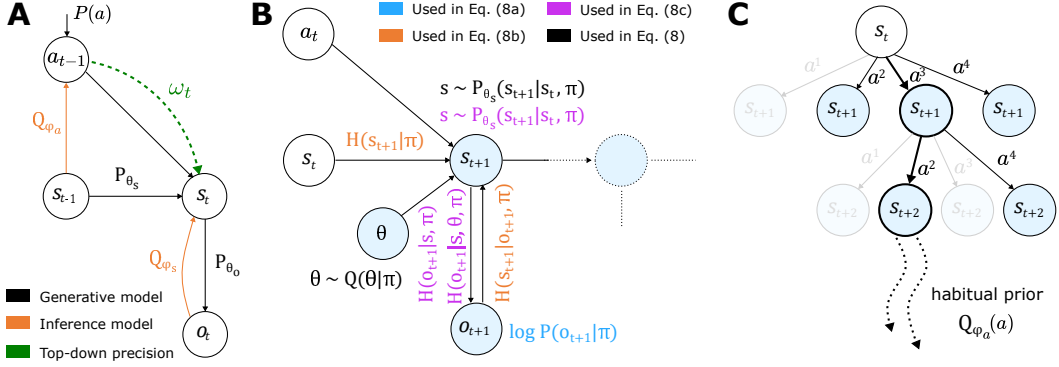}
\centering
\caption{\textbf{A}: Schematic of model architecture and networks used during the learning process. Black arrows represent the generative model ($P$), orange arrows the recognition model ($Q$), and the green arrow the top-down attention ($\bm\omega_t$). \textbf{B}: Relevant quantities for the calculation of EFE $G$, computed by simulating the future using the generative model and ancestral sampling. Where appropriate, expectations are taken with a single MC sample. \textbf{C}: MCTS scheme used for planning and acting, using the habitual network to selectively explore new tree branches.}
\label{fig:architecture}
\end{figure}

\subsection{Calculating variational and expected free energy}
\label{sec:calculating}

First, we extend the probabilistic graphical model (as defined in Sec.~\ref{activeinference}) to include the action sequences $\pi$ and factorize the model based on Fig.~\subref{fig:architecture}{A}. We then exploit standard variational inference machinery to calculate the free energy for each time-step $t$ as:
\begin{align}
\begin{split}\label{eq:Ft}
F_t = &-\E{Q_{\phi_s}(s_t)}\big[ \log P_{\theta_o}(o_t|s_t) \big] + \KL{Q_{\phi_s}(s_t)}{P_{\theta_s}(s_t|s_{t-1},a_{t-1})} \\
& + \mathop{\mathbb{E}_{Q_{\phi_s}(s_t)}}\left[ \KL{Q_{\phi_a}(a_t)}{P(a_t)} \right] ~ ,
\end{split}
\end{align}
where
\begin{equation}\label{eq:Pa}
P(a)=\sum_{\pi:a_1=a}{P(\pi)}
\end{equation}
is the summed probability of all policies that begin with action $a$.

We assume that $s_t$ is normally distributed and $o_t$ is Bernoulli distributed, with all parameters given by a neural network, parameterized by $\theta_o$, $\theta_s$, and $\phi_s$ for the observation, transition, and encoder models, respectively (see Sec.~\ref{sec:methods:habit} for details about $Q_{\phi_a}$). With this assumption, all the terms here are standard log-likelihood and KL terms easy to compute for Gaussian and Bernoulli distributions. The expectations over $Q_{\phi_s}(s_t)$ are taken via MC sampling, using a single sample from the encoder. 

Next, we consider EFE. At time-step $t$ and for a time horizon up to time $T$, EFE is defined as~\cite{friston2017active}:
\begin{align}
    G(\pi) = \sum_{\tau=t}^T G(\pi, \tau) = \sum_{\tau=t}^T \E{\tilde{Q}} \left[ \log Q(s_\tau, \theta | \pi) - \log \tilde{P}(o_\tau, s_\tau, \theta | \pi) \right] ~ ,
\end{align}
where $\tilde{Q} = Q(o_\tau, s_\tau, \theta | \pi) =Q(\theta|\pi)Q(s_{\tau}|\theta,\pi)Q(o_{\tau}|s_{\tau},\theta,\pi)$ and $\tilde{P}(o_\tau, s_\tau, \theta | \pi) = P(o_\tau | \pi) Q(s_\tau | o_\tau) P(\theta | s_\tau, o_\tau)$. Following Schwartenbeck \emph{et al.}~\cite{schwartenbeck2019computational}, the EFE of a single time instance $\tau$ can be further decomposed as
\begin{subequations}\label{eq:G3}
\begin{align}\label{eq:G3a}
G(\pi,\tau) = &-\E{\tilde{Q}}\big[ \log P(o_{\tau}|\pi) \big] \\\label{eq:G3b}
&+\E{\tilde{Q}}\big[ \log Q(s_{\tau}|\pi) - \log P(s_{\tau}|o_{\tau},\pi) \big]\\\label{eq:G3c}
&+\E{\tilde{Q}}\big[ \log Q(\theta|s_{\tau},\pi) - \log P(\theta|s_{\tau},o_{\tau},\pi) \big] ~ .
\end{align}
\end{subequations}
Interestingly, each term constitutes a conceptually meaningful expression. The term \eqref{eq:G3a} corresponds to the likelihood assigned to the desired observations $o_{\tau}$, and plays an analogous role to the notion of reward in the reinforcement learning (RL) literature~\cite{sutton2018reinforcement}. The term \eqref{eq:G3b} corresponds to the mutual information between the agent's beliefs about its latent representation of the world, before and after making a new observation, and hence, it reflects a motivation to explore areas of the environment that resolve state uncertainty. Similarly, the term \eqref{eq:G3c} describes the tendency of active inference agents to reduce their uncertainty about model parameters via new observations and is usually referred to in the literature as active learning~\cite{learningpaper}, novelty, or curiosity~\cite{schwartenbeck2019computational}.

However, two of the three terms that constitute EFE cannot be easily computed as written in Eq.~\eqref{eq:G3}. To make computation practical, we will re-arrange these expressions and make further use of MC sampling to render these expressions tractable and re-write Eq.~\eqref{eq:G3} as
\begin{subequations}\label{eq:G5}
\begin{align}\label{eq:G5a}
G(\pi,\tau) = &-\mathop{\mathbb{E}_{Q(\theta|\pi)Q(s_{\tau}|\theta,\pi)Q(o_{\tau}|s_{\tau},\theta,\pi)}}\big[ \log P(o_{\tau}|\pi) \big] \\\label{eq:G5b}
&+\mathop{\mathbb{E}_{Q(\theta|\pi)}} \big[ \E{Q(o_{\tau}|\theta,\pi)} H(s_{\tau}|o_{\tau}, \pi) - H(s_\tau | \pi) \big]\\\label{eq:G5c}
&+\mathop{\mathbb{E}_{Q(\theta|\pi)Q(s_{\tau}|\theta,\pi)}}H(o_{\tau}|s_{\tau},\theta,\pi) - \mathop{\mathbb{E}_{Q(s_{\tau}|\pi)}} H(o_{\tau}|s_{\tau},\pi) ~ ,
\end{align}
\end{subequations}
where these expressions can be calculated from the deep neural network illustrated in Fig.~\subref{fig:architecture}{B}. The derivation of Eq.~(\ref{eq:G5}) can be found in the supplementary material. To calculate the terms (\ref{eq:G5a}) and (\ref{eq:G5b}), we sample $\theta$, $s_{\tau}$ and $o_{\tau}$ sequentially (through ancestral sampling) and then $o_{\tau}$ is compared with the prior distribution $\log P(o_{\tau}|\pi)$. The parameters of the neural network $\theta$ are sampled from $Q(\theta)$ using the MC dropout technique~\cite{Gal:2016:Dropout}. Similarly, to calculate the expectation of $H(o_{\tau}|s_{\tau},\pi)$, the same drawn $\theta$ is used again and $s_{\tau}$ is re-sampled for $N$ times while, for $H(o_{\tau}|s_{\tau},\theta,\pi)$, the set of parameters $\theta$ is also re-sampled $N$ times. Finally, all entropies can be computed using the standard formulas for multivariate Gaussian and Bernoulli distributions.

\subsection{Action selection and the habitual network}
\label{sec:methods:habit}

In active inference, agents choose an action given by their EFE. In particular, any given action is selected with a probability proportional to the accumulated negative EFE of the corresponding policies $G(\pi)$ (see Eq.~\eqref{eq:G1} and Ref.~\cite{parr2019generalised}). However, computing $G$ across all policies is costly since it involves making an exponentially-increasing number of predictions for $T$-steps into the future, and computing all the terms in Eq.~\eqref{eq:G5}. To solve this problem, we employ two methods operating in tandem. First, we employ standard MCTS~\cite{coulom2006efficient,browne2012survey,silver2017mastering}, a search algorithm in which different potential future trajectories of states are explored in the form of a search tree (Fig.~\subref{fig:architecture}{C}), giving emphasis to the most likely future trajectories. This algorithm is used to calculate the distribution over actions $P(a_t)$, defined in Eq.~\eqref{eq:Pa}, and control the agent's final decisions. Second, we make use of amortized inference through a habitual neural network that directly approximates the distribution over actions, which we parameterize by $\phi_a$ and denote $Q_{\phi_a}(a_t)$ -- similarly to Refs.~\cite{piche2018probabilistic,tschantz2020control,marinoinference}. In essence, $Q_{\phi_a}(a_t)$ acts as a variational posterior that approximates $P(a_t|s_t)$, with a prior $P(a_t)$, calculated by MCTS (see Fig.~\subref{fig:architecture}{A}). During learning, this network is trained to reproduce the last executed action $a_{t-1}$ (selected by sampling $P(a_t)$) using the last state $s_{t-1}$. Since both tasks used in this paper (Sec.~\ref{sec:results}) have discrete action spaces $\mathcal{A}$, we define $Q_{\phi_a}(a_t)$ as a neural network with parameters $\phi_a$ and $|\mathcal{A}|$ softmax output units.

During the MCTS process, the agent generates a weighted search tree iteratively that is later sampled during action selection. In each single MCTS loop, one plausible state-action trajectory $(s_t,a_t,s_{t+1},a_{t+1},...,s_{\tau},a_{\tau})$ -- starting from the present time-step $t$ -- is calculated. For states that are explored for the first time, the distribution $P_{\theta_s}(s_{t+1}|s_t,a_t)$ is used. States that have been explored are stored in the \textit{buffer} search tree and accessed during later loops of the same planning process. The weights of the search tree $\tilde{G}(s_t,a_t)$ represent the agent's best estimation for EFE after taking action $a_t$ from state $s_t$. An upper confidence bound for $G(s_t,a_t)$ is defined as
\begin{equation} \label{eq:MCTS:U}
U(s_t,a_t)=\tilde{G}(s_t,a_t) + c_{\text{explore}} \cdot Q_{\phi_a}(a_t|s_t) \cdot \frac{1}{1+N(a_t,s_t)} ~ ,
\end{equation}

where $N(a_t,s_t)$ is the number of times that $a_t$ was explored from state $s_t$, and $c_{\text{explore}}$ a hyper-parameter that controls exploration. In each round, the EFE of the newly-explored parts of the trajectory is calculated and back-propagated to all visited nodes of the search tree. Additionally, actions are sampled in two ways. Actions from states that have been explored are sampled from $\sigma(U(a_t,s_t))$ while actions from new states are sampled from $Q_{\phi_a}(a_t)$. 

Finally, the actions that assemble the selected policy are drawn from $P(a_t) = \frac{N(a_t,s_t)}{\sum_j{N(a_{j,t},s_t)}}$. In our implementation, the planning loop stops if either the process has identified a clear option (i.e. if $\max{P(a_t)}- 1/|\mathcal{A}| > T_{dec}$) or the maximum number of allowed loops has been reached. 

Through the combination of the approximation $Q_{\phi_a}(a_t)$ and the MCTS, our agent has at its disposal two methods of action selection. We refer to $Q_{\phi_a}(a_t)$ as the \emph{habitual} network, as it corresponds to a form of fast decision-making, quickly evaluating and selecting a action; in contrast with the more \emph{deliberative} system that includes future imagination via MC tree traversals~\cite{van2012information}.

\subsection{State precision and top-down attention}
\label{sec:methods:precision}

One of the key elements of our framework is the state transition model $P_{\theta_s}(s_t | s_{t-1}, a_{t-1})$, that belongs to the agent's generative model. In our implementation, we take $s_t \sim \mathcal{N}(\mu, \sigma^2/{\bm\omega_t})$, where the multidimensional $\mu$ and $\sigma$ come from the linear and softplus units (respectively) of a neural network with parameters $\theta_s$ applied to $s_{t-1}$, and, importantly, $\bm\omega_t$ is a scalar \emph{precision factor} (c.f. Fig.~\subref{fig:architecture}{A}) modulating the uncertainty on the agent's estimate of the hidden state of the environment~\cite{Parr:2017:Uncertainty}. We model the precision factor as a simple logistic function of the belief update about the agent's current policy,
\begin{align}
\bm\omega_t = \frac{\alpha}{1 + e^{- \frac{b-D_{t-1}}{c} }} + d ~ ,
\label{eq:precision}%
\end{align}
where $D_t = \KL{Q_{\phi_a}(a_t)}{P(a_t)}$ and $\{\alpha,b,c,d\}$ are fixed hyper-parameters. Note that $\bm\omega_t$ is a monotonically decreasing function of $D_{t-1}$, such that when the posterior belief about the current policy is similar to the prior, precision is high.

In cognitive terms, $\bm\omega_t$ can be thought of as a means of \emph{top-down attention}~\cite{byers2012topdown}, that regulates which transitions should be learnt in detail and which can be learnt less precisely. This attention mechanism acts as a form of resource allocation: if $\KL{Q_{\phi_a}(a_t)}{P(a_t)}$ is high, then a habit has not yet been formed, reflecting a generic lack of knowledge. Therefore, the precision of the prior $P_{\theta_s}(s_t|s_{t-1},a_{t-1})$ (i.e., the belief about the current state before a new observation $o_t$ has been received) is low, and less effort is spent learning $Q_{\phi_s}(s_t)$.

In practice, the effect of $\bm\omega_t$ is to \emph{incentivize disentanglement} in the latent state representation $s_t$ -- the precision factor $\bm\omega_t$ is somewhat analogous to the $\beta$ parameter in $\beta$-VAE~\cite{Higgins:2017:betaVAE}, effectively pushing the state encoder $Q_{\phi_s}(s_t)$ to have independent dimensions (since $P_{ \theta_s}(s_t|s_{t-1},a_{t-1})$ has a diagonal covariance matrix).\footnote{In essence, for the parameter ranges of interest $\bm\omega_t$ induces a near-linear monotonic increase in $D_{\text{KL}}$, akin to the linear increase induced by $\beta$ in $\beta$-VAE.} As training progresses and the habitual network becomes a better approximation of $P(a_t)$, $\bm\omega_t$ is gradually increased, implementing a natural form of precision annealing.

\section{Results}\label{sec:results}
First, we present the two environments that were used to validate our agent's performance. 

\begin{figure}[h]
\includegraphics[width=0.9\linewidth]{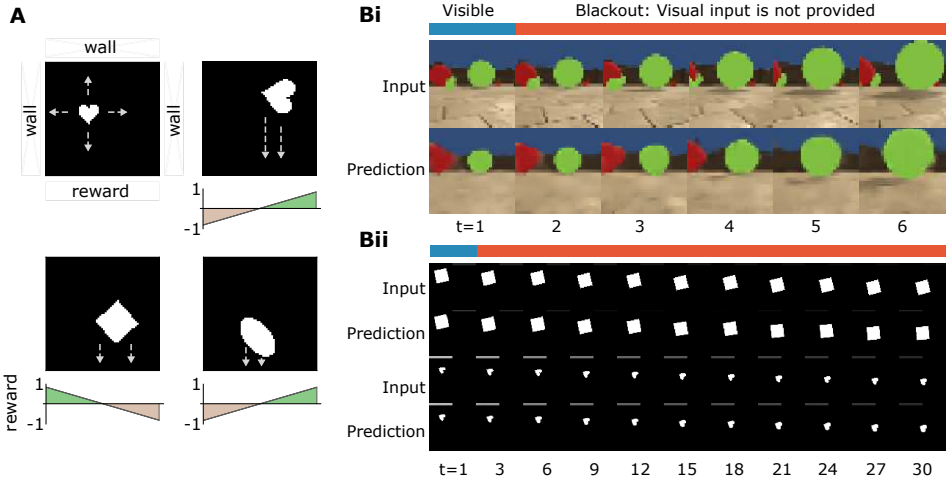}
\centering
\caption{\textbf{A:} The proposed \textit{object sorting} task based on the dSprites dataset. The agent can perform 4 actions; changing the position of the object in both axis. Reward is received if an object crosses the bottom boarder and differs for the 3 object shapes. \textbf{B:} Prediction of the visual observations under motion if input is hidden in both (\textbf{i}) AnimalAI and (\textbf{ii}) dynamic dSprites environments. }
\label{fig:environments}
\end{figure}

\paragraph{Dynamic dSprites} We defined a simple 2D environment based on the dSprites dataset~\cite{dsprites17,Higgins:2017:betaVAE}. This was used to $i)$ quantify the agent's behavior against ground truth state-spaces and $ii)$ evaluate the agent's ability to disentangle state representations. This is feasible as the dSprites data is designed for characterizing disentanglement, using a set of interpretable, independent ground-truth latent factors. In this task, which we call \textit{object sorting}, the agent controls the position of the object via $4$ different actions (right, left, up or down) and is required to sort single objects based on their shape (a latent factor). The agent receives reward when it moves the object across the bottom border, and the reward value depends on the shape and location as depicted in Fig.~\subref{fig:environments}{A}. For the results presented in Section~\ref{sec:results}, the agent was trained in an on-policy fashion, with a batch size of $100$. 

\paragraph{Animal-AI} We used a variation of \textit{`preferences'} task from the Animal-AI environment~\cite{Crosby:2019:animalai}. The complexity of this, partially observable, 3D environment is the ideal test-bed for showcasing the agent's reward-directed exploration of the environment, whilst avoiding negative reward or getting stuck in corners. In addition, to test the agent's ability to rely on its internal model, we used a `\textit{lights-off}' variant of this task, with temporary suspension of visual input at any given time-step with probability $R$. For the results presented in Section~\ref{sec:results}, the agent was trained in an off-policy fashion due to computational constraints. The training data for this was created using a simple rule: move in the direction of the greenest pixels.

In the experiments that follow, we encode the actual reward from both environments as the prior distribution of future expected observations $\log P(o_{\tau}|\pi)$ or, in active inference terms, the expected outcomes. This is appropriate because the active inference formulation does not differentiate reward from other types of observations, but it rather defines certain (future) observations (e.g. green color in Animal-AI) as more desirable given a task. Therefore, in practice, rewards can be encoded as observations with higher prior probability using $\log P(o_\tau|\pi)$.

We optimized the networks using ADAM \cite{kingma2014adam}, with loss given in Eq.~\eqref{eq:Ft} and an extra regularization term $\KL{Q_{\phi_s}(s_t)}{N(0,1)}$. The explicit training procedure is detailed in the supplementary material. The complete source-code, data, and pre-trained agents, is available on GitHub \texttt{(\href{https://github.com/zfountas/deep-active-inference-mc}{https://github.com/zfountas/deep-active-inference-mc})}.

\subsection{Learning environment dynamics and task performance}

\begin{figure}[h]
\includegraphics[width=0.88\linewidth]{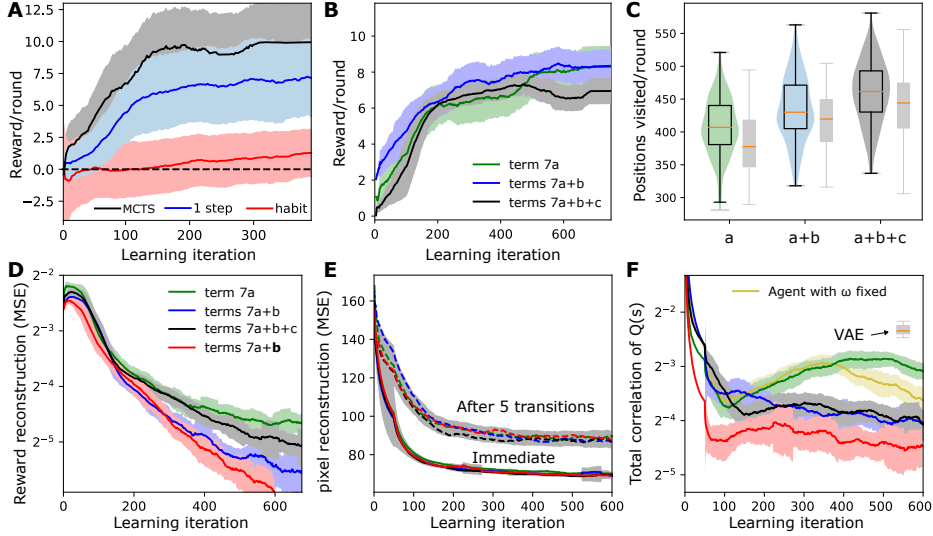}
\centering
\caption{Agent's performance during on-policy training in the \textit{object sorting} task. \textbf{A:} Comparison of different action selection strategies for the agent driven by the full Eq. (\ref{eq:G5}).
\textbf{B-C:} Comparison of agents driven by different functionals, limited to state estimations of a single step into the future. In C, the violin plots represent behavior driven by $P(a_t)$ (the planner) and the gray box plots driven by the habitual network $Q_{\phi_a}(a_t)$. 
\textbf{D-F:} Reconstruction loss and total correlation during learning for 4 different functionals. In A, B and D-F, the shaded areas represent the standard deviation. }
\label{fig:results:dsprites}
\end{figure}

We initially show -- through a simple visual demonstration (Fig.~\ref{fig:environments}B) -- that agents learn the environment dynamics with or without consistent visual input for both dynamic dSprites and AnimalAI. This is further investigated, for the dynamic dSprites, by evaluating task performance (Fig.~\subref{fig:results:dsprites}{A-C}), as well as reconstruction loss for both predicted visual input and reward (Fig.~\subref{fig:results:dsprites}{D-E}) during training. 

To explore the effect of using different EFE functionals on behavior, we trained and compared active inference agents under three different formulations, all of which used the implicit reward function $\log P(o_\tau)$, against a baseline reward-maximizing agent. These include $i$) beliefs about the latent states (i.e., terms a,b from Eq.~\ref{eq:G3}), $ii$) beliefs about both the latent states and model parameters (i.e., complete Eq.~\ref{eq:G3}) and $iii$) beliefs about the latent states, with a down-weighted reward signal. We found that, although all agents exhibit similar performance in collecting rewards (Fig.~\subref{fig:results:dsprites}{B}), active inference agents have a clear tendency to explore the environment (Fig.~\subref{fig:results:dsprites}{C}). Interestingly, our results also demonstrate that all three formulations are better at reconstructing the expected reward, in comparison to a reward-maximizing baseline (Fig.~\subref{fig:results:dsprites}{D}). Additionally, our agents are capable of reconstructing the current observation, as well as predicting 5 time-steps into the future, for all formulations of EFE, with similar loss with the baseline (Fig.~\subref{fig:results:dsprites}{E}).

\subsection{Disentanglement and transition learning}

Disentanglement of latent spaces leads to lower dimensional temporal dynamics that are easier to predict~\cite{Hsieh:2018:nips}. Thus, generating a disentangled latent space $s$ can be beneficial for learning the parameters of the transition function $P_{ \theta_s}(s_{t+1} | s_t, a_t)$. Due to the similarity between the precision term $\bm\omega_t$ and the hyper-parameter $\beta$ in $\beta$-variational autoencoders (VAEs)~\cite{Higgins:2017:betaVAE} discussed in Sec.~\ref{sec:methods:precision}, we hypothesized that $\bm\omega_t$ could play an important role in regulating transition learning.
To explore this hypothesis, we compared the total correlation (as a metric for disentanglement~\cite{Kim:2018:factorvae}) of latent state beliefs between $i)$ agents that have been trained with the different EFE functionals, $ii)$ the baseline (reward-maximizing) agent, $iii)$ an agent trained without top-down attention (although the average value of $\bm\omega_t$ was maintained), as well as $iv)$ a simple VAE that received the same visual inputs. As seen in Fig.~\subref{fig:results:dsprites}{F}, all active inference agents using $\bm\omega_t$ generated structures with significantly more disentanglement (see traversals in supp. material). Indeed, the performance ranking here is the same as in Fig.~\subref{fig:results:dsprites}{D}, pointing to disentanglement as a possible reason for the performance difference in predicting rewards.

\subsection{Navigation and planning in reward-based tasks}

\begin{figure}[h!]
\includegraphics[width=\linewidth]{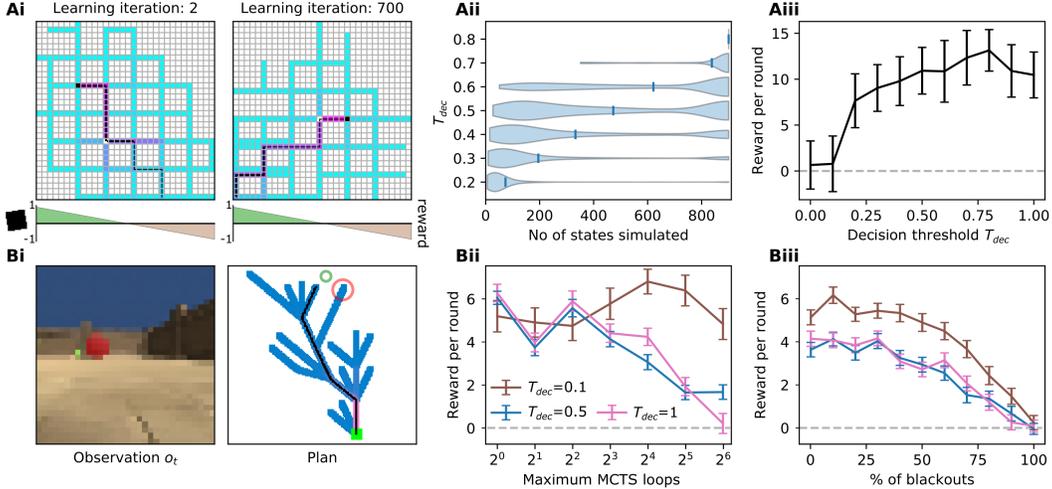}
\centering
\caption{Agent's planning performance. \textbf{A:} Dynamic dSprites. \textbf{i)} Example planned trajectory plots with number of visits per state (blue-pink color map) and the selected policy (black lines). \textbf{ii)} The effect of decision threshold $T_{\text{dec}}$ on the number of simulated states and \textbf{iii)} the agent's performance. \textbf{B:} Animal-AI. \textbf{i)} Same as in A. \textbf{ii)} System performance over hyper-parameters and  \textbf{iii)} in the \textit{lights-off} task. Error bars in Aiii denote standard deviation and in B standard error of the mean.}
\label{fig:planning}
\end{figure}

The training process in the dynamic dSprites environment revealed two types of behavior. Initially, we see epistemic exploration (i.e., curiosity), that is overtaken by reward seeking (i.e., goal-directed behavior) once the agent is reasonably confident about the environment. An example of this can be seen in the left trajectory plot in Fig.~\subref{fig:planning}{Ai}, where the untrained agent -- with no concept of reward -- deliberates between multiple options and chooses the path that enables it to quickly move to the next round. The same agent, after $700$ learning iterations, can now optimally plan where to move the current object, in order to maximize potential reward, $\log P(o_{\tau}|\pi)$. We next investigated the sensitivity when deciding, by changing the threshold $T_{dec}$. We see that changing the threshold has clear implications for the distribution of explored alternative trajectories i.e., number of simulated states (Fig.~\subref{fig:planning}{Aii}). This plays an important role in the performance, with maximum performance found at $T_{dec} \approx 0.8$ (Fig.~\subref{fig:planning}{Aiii}).

Agents trained in the Animal-AI environment also exhibit interesting (and intelligent) behavior. Here, the agent is able to make complex plans, by avoiding obstacles with negative reward and approaching expected outcomes (red and green objects respectively, Fig.~\subref{fig:planning}{Bi}). 
Maximum performance can be found for $16$ MCTS loops and  
$T_{dec} \approx 0.1$ (Fig.~\subref{fig:planning}{Bii}; details in the supplementary material). When deployed in \textit{lights-off} experiments, the agent can successfully maintain an accurate representation of the world state and simulate future plans despite temporary suspension of visual input (Fig.~\subref{fig:environments}{B}). This is particularly interesting because $P_{\theta_s}(s_{t+1} | s_t, a_t)$ is defined as a feed-forward network, without the ability to maintain memory of states before $t$. As expected, the agent's ability to operate in this set-up becomes progressively worse the longer the visual input is removed, while shorter decision thresholds are found to preserve performance longer (Fig.~\subref{fig:planning}{Biii}).

\subsection{Comparison with model-free reinforcement learning agents}
\label{sec:results:rl}

\begin{figure}[h]
\includegraphics[width=0.55\linewidth]{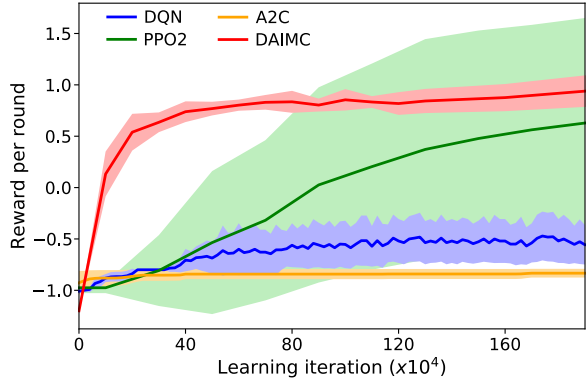}
\centering
\caption{Comparison of the performance of our agent (DAIMC) with DQN, A2C and PPO2. The bold line represents the mean, and shaded areas the standard deviation over multiple training runs.}
\label{fig:results:modelfree}
\end{figure}

To assess the performance of the active inference agent with respect to baseline (model-free) RL agents, we employed OpenAI's \texttt{baselines}~\cite{baselines} repository to train DQN~\cite{Mnih:2015:dqn}, A2C~\cite{Mnih:2016:A2C} and PPO2~\cite{Schulman:2017:PPO} agents on the Animal-AI environment. The resulting comparison is shown in Fig.~\ref{fig:results:modelfree}. Our experiments highlight that, given the same number of training episodes (2M learning iterations), the active inference agent performs considerably better than DQN and A2C, and is comparable to PPO2 (all baselines were trained with default settings). In addition, note that of the two best-performing agents (DAIMC and PPO2), DAIMC has substantially less variance across training runs, indicating a more stable learning process. Nonetheless, these comparisons should be treated as a way of illustrating the applicability of the active inference agent operating in complex environments, and not as a thorough benchmark of performance gains against state-of-the-art RL agents.  

\section{Concluding Remarks}
\label{sec:conclusion}

The attractiveness of active inference inherits from the biological plausibility of the framework~\cite{friston2017active,isomura2018vitro,adams2013predictions}. 
Accordingly, we focused on scaling-up active inference inspired by neurobiological structure and function that supports intelligence. This is reflected in the hierarchical generative model, where the higher-level policy network contextualizes lower-level state representations. This speaks to a separation of temporal scales afforded by cortical hierarchies in the brain and provides a flexible framework to develop biologically-inspired intelligent agents.

We introduced MCTS for tackling planning problems with vast search spaces \cite{coulom2006efficient,kocsis2006bandit,browne2012survey,guo2014deep,silver2016mastering}. This approach builds upon {\c{C}}atal \textit{et al.}'s~\cite{ccatal2020learning} deep active inference proposal, to use tree search to recursively re-evaluate EFE for each policy, but is computationally more efficient. Additionally, using MCTS offers an \textit{Occam's window} for policy pruning; that is, we stop evaluating a policy path if its EFE becomes much higher than a particular upper confidence bound. This pruning drastically reduces the number of paths one has to evaluate. It is also consistent with biological planning, where agents adopt brute force exploration of possible paths in a decision tree, up to a resource-limited finite depth~\cite{snider2015prospective}. This could be due to imprecise evidence about different future trajectories~\cite{solway2015evidence} where environmental constraints subvert evaluation accuracy~\cite{van2017computational,holding1989counting} or alleviate computational load~\cite{huys2012bonsai}. Previous work addressing the depth of possible future trajectories in human subjects under changing conditions shows that both increased cognitive load~\cite{holding1989counting} and time constraints~\cite{burns2004effects,van2007effects,van2017computational} reduce search depth. Huys \textit{et al.}~\cite{huys2012bonsai} highlighted that in tasks involving alleviated computational load, subjects might evaluate only subsets of decision trees. This is consistent with our experiments as the agent selects to evaluate only particular trajectories based on their prior probability to occur.

We have shown that the precision factor, $\bm\omega_t$, can be used to incorporate uncertainty over the prior and enhances disentanglement by encouraging statistical independence between features~\cite{Mathieu:2018,Kim:2019,hadi2007attention,mott2019towards}. This is precisely why it has been associated with attention~\cite{parr2018perceptual}; a signal that shapes uncertainty~\cite{dayan2000learning}. Attention enables flexible modulation of neural activity that allows behaviorally relevant sensory data to be processed more efficiently~\cite{baluch2011mechanisms,sasaki2010advances,byers2012topdown}. The neural realizations of this have been linked with neuromodulatory systems, e.g., cholinergic and noradrenergic~\cite{posner1990attention,dayan2001nips,gu2002neuromodulatory,Yu:2005:Uncertainty,moran2013free}. In active inference, they have been associated specifically with noradrenaline for modulating uncertainty about state transitions~\cite{Parr:2017:Uncertainty}, noradrenergic modulation of visual attention~\cite{parr2019computational} and dopamine for policy selection~\cite{friston2017active,parr2019computational}.

An important piece of future work is to more thoroughly compare the performance of DAIMC agent to reward-maximizing agents. That is, if the specific goal is to maximize reward, then it is not clear whether deep active inference (i.e., full specification of EFE) has significant performance benefits over simpler reward-seeking agents (i.e., using only Eq.~\ref{eq:G3}a) or other model-based RL agents \cite{fellows2019virel,levine2018reinforcement} (c.f. Sec.~\ref{sec:results:rl}). We emphasize, however, that the primary purpose of the active inference framework is to serve as a model for biological cognition, and not as an optimal solution for reward-based tasks. Therefore, we have deliberately not focused on benchmarking performance gains against state-of-the-art RL agents, although we hypothesize that insights from active inference could prove useful in complex environments where either reward maximization isn't the objective, or in instances where direct reward maximization leads to sub-optimal performance.

There are several extensions that can be explored, such as testing whether performance would increase with more complex, larger neural networks, e.g., using LSTMs to model state transitions. One could also assess if including episodic memory would finesse EFE evaluation over a longer time horizon, without increasing computational complexity. Future work should also test how performance shifts if the objective of the task changes. Lastly, it might be neurobiologically interesting to see whether the generated disentangled latent structures are apt for understanding functional segregation in the brain.

\section{Broader impact}
Our deep active inference agent – equipped with MC methods – provides a flexible framework that may help gain new insights into the brain by simulating realistic, biologically-inspired intelligent agents. General contributions of this framework include helping bridge the gap between cognitive science and deep learning and  providing an architecture that would allow psychologists to run more realistic experiments probing human behavior. Specifically, we hope that simulating this agent will allow us use the neural network gradients to make predictions about the underlying physiology associated with behaviors of interest and formulate appropriate hypothesis. We believe this architecture may also help elucidate complex structure-function relationships in cognitive systems through manipulation of priors (under the complete class theorem). This would make it a viable (scaled-up) framework for understanding how brain damage (introduced in the generative model by changing the priors) can affect cognitive function, previously explored in discrete-state formulations of active inference \cite{parr2019computational,parr2018computational}.

A potential (future) drawback is that this model could be used to exploit people's inherent cognitive biases, and as such could potentially be used by bad actors trying to model (and then profit from) human behavior.

\section{Acknowledgements}
The authors would like to thank Sultan Kenjeyev for his valuable contributions and comments on early versions of the model presented in the current manuscript and Emotech team for the great support throughout the project. NS was funded by the Medical Research Council (MR/S502522/1). PM and KJF were funded by the Wellcome Trust (Ref: 210920/Z/18/Z - PM; Ref: 088130/Z/09/Z - KJF).

\bibliography{paper}

\begin{thebibliography}{70}
\providecommand{\natexlab}[1]{#1}
\providecommand{\url}[1]{\texttt{#1}}
\expandafter\ifx\csname urlstyle\endcsname\relax
  \providecommand{\doi}[1]{doi: #1}\else
  \providecommand{\doi}{doi: \begingroup \urlstyle{rm}\Url}\fi

\bibitem[Friston(2010)]{friston2010free}
Karl~J Friston.
\newblock The free-energy principle: A unified brain theory?
\newblock \emph{Nature Reviews Neuroscience}, 11\penalty0 (2):\penalty0
  127--138, 2010.

\bibitem[Friston(2019)]{friston2019free}
Karl~J Friston.
\newblock A free energy principle for a particular physics.
\newblock \emph{arXiv preprint arXiv:1906.10184}, 2019.

\bibitem[Friston et~al.(2016)Friston, FitzGerald, Rigoli, Schwartenbeck,
  O'Doherty, and Pezzulo]{learningpaper}
Karl~J Friston, Thomas. FitzGerald, Francesco Rigoli, Philipp Schwartenbeck,
  John O'Doherty, and Giovanni Pezzulo.
\newblock Active inference and learning.
\newblock \emph{Neuroscience $\&$ Biobehavioral Reviews}, 68:\penalty0 862--79,
  2016.

\bibitem[Friston et~al.(2017{\natexlab{a}})Friston, FitzGerald, Rigoli,
  Schwartenbeck, and Pezzulo]{friston2017active}
Karl~J Friston, Thomas FitzGerald, Francesco Rigoli, Philipp Schwartenbeck, and
  Giovanni Pezzulo.
\newblock Active inference: {A} process theory.
\newblock \emph{Neural Computation}, 29\penalty0 (1):\penalty0 1--49,
  2017{\natexlab{a}}.

\bibitem[Friston et~al.(2017{\natexlab{b}})Friston, Parr, and
  de~Vries]{friston2017graphical}
Karl~J Friston, Thomas Parr, and Bert de~Vries.
\newblock The graphical brain: {Belief} propagation and active inference.
\newblock \emph{Network Neuroscience}, 1\penalty0 (4):\penalty0 381--414,
  2017{\natexlab{b}}.

\bibitem[Pezzulo et~al.(2018)Pezzulo, Rigoli, and
  Friston]{pezzulo2018hierarchical}
Giovanni Pezzulo, Francesco Rigoli, and Karl~J Friston.
\newblock Hierarchical active inference: A theory of motivated control.
\newblock \emph{Trends in Cognitive Sciences}, 22\penalty0 (4):\penalty0
  294--306, 2018.

\bibitem[Da~Costa et~al.(2020)Da~Costa, Parr, Sajid, Veselic, Neacsu, and
  Friston]{da2020active}
Lancelot Da~Costa, Thomas Parr, Noor Sajid, Sebastijan Veselic, Victorita
  Neacsu, and Karl Friston.
\newblock Active inference on discrete state-spaces: {A} synthesis.
\newblock \emph{arXiv preprint arXiv:2001.07203}, 2020.

\bibitem[Parr and Friston(2017)]{Parr:2017:Uncertainty}
Thomas Parr and Karl~J Friston.
\newblock Uncertainty, epistemics and active inference.
\newblock \emph{Journal of The Royal Society Interface}, 14\penalty0
  (136):\penalty0 20170376, 2017.

\bibitem[Friston et~al.(2018)Friston, Rosch, Parr, Price, and
  Bowman]{deeptemporal}
Karl~J Friston, Richard Rosch, Thomas Parr, Cathy Price, and Howard Bowman.
\newblock Deep temporal models and active inference.
\newblock \emph{Neuroscience and Biobehavioral Reviews}, 90:\penalty0
  486—501, 2018.

\bibitem[Sajid et~al.(2019)Sajid, Ball, and Friston]{sajid2019demystifying}
Noor Sajid, Philip~J Ball, and Karl~J Friston.
\newblock Active inference: {Demystified} and compared.
\newblock \emph{arXiv preprint arXiv:1909.10863}, 2019.

\bibitem[Hesp et~al.(2019)Hesp, Smith, Allen, Friston, and
  Ramstead]{hesp2019deeply}
Casper Hesp, Ryan Smith, Micah Allen, Karl~J Friston, and Maxwell Ramstead.
\newblock Deeply felt affect: {The} emergence of valence in deep active
  inference.
\newblock \emph{PsyArXiv}, 2019.

\bibitem[Cullen et~al.(2018)Cullen, Davey, Friston, and Moran]{Cullen}
Maell Cullen, Ben Davey, Karl~J Friston, and Rosalyn~J. Moran.
\newblock Active inference in {OpenAI Gym: A} paradigm for computational
  investigations into psychiatric illness.
\newblock \emph{Biological Psychiatry: Cognitive Neuroscience and
  Neuroimaging}, 3\penalty0 (9):\penalty0 809--818, 2018.

\bibitem[Friston et~al.(2009)Friston, Daunizeau, and Kiebel]{active_rl}
Karl~J Friston, Jean Daunizeau, and Stefan~J Kiebel.
\newblock Reinforcement learning or active inference?
\newblock \emph{PLoS ONE}, 4\penalty0 (7):\penalty0 e6421, 2009.

\bibitem[Ueltzh{\"o}ffer(2018)]{ueltzhoffer2018deep}
Kai Ueltzh{\"o}ffer.
\newblock Deep active inference.
\newblock \emph{Biological Cybernetics}, 112\penalty0 (6):\penalty0 547--573,
  2018.

\bibitem[{\c{C}}atal et~al.(2019){\c{C}}atal, Nauta, Verbelen, Simoens, and
  Dhoedt]{ccatal2019bayesian}
Ozan {\c{C}}atal, Johannes Nauta, Tim Verbelen, Pieter Simoens, and Bart
  Dhoedt.
\newblock Bayesian policy selection using active inference.
\newblock \emph{arXiv preprint arXiv:1904.08149}, 2019.

\bibitem[Tschantz et~al.(2019)Tschantz, Baltieri, Seth, Buckley,
  et~al.]{tschantz2019scaling}
Alexander Tschantz, Manuel Baltieri, Anil Seth, Christopher~L Buckley, et~al.
\newblock Scaling active inference.
\newblock \emph{arXiv preprint arXiv:1911.10601}, 2019.

\bibitem[Millidge(2020)]{millidge2020deep}
Beren Millidge.
\newblock Deep active inference as variational policy gradients.
\newblock \emph{Journal of Mathematical Psychology}, 96:\penalty0 102348, 2020.

\bibitem[Blei et~al.(2017)Blei, Kucukelbir, and McAuliffe]{blei2017variational}
David~M Blei, Alp Kucukelbir, and Jon~D McAuliffe.
\newblock Variational inference: A review for statisticians.
\newblock \emph{Journal of the American Statistical Association}, 112\penalty0
  (518):\penalty0 859--877, 2017.

\bibitem[Parr and Friston(2019)]{parr2019generalised}
Thomas Parr and Karl~J Friston.
\newblock Generalised free energy and active inference.
\newblock \emph{Biological Cybernetics}, 113\penalty0 (5-6):\penalty0 495--513,
  2019.

\bibitem[Schwartenbeck et~al.(2019)Schwartenbeck, Passecker, Hauser,
  FitzGerald, Kronbichler, and Friston]{schwartenbeck2019computational}
Philipp Schwartenbeck, Johannes Passecker, Tobias~U Hauser, Thomas~HB
  FitzGerald, Martin Kronbichler, and Karl~J Friston.
\newblock Computational mechanisms of curiosity and goal-directed exploration.
\newblock \emph{eLife}, 8:\penalty0 e41703, 2019.

\bibitem[Sutton and Barto(2018)]{sutton2018reinforcement}
Richard~S Sutton and Andrew~G Barto.
\newblock \emph{Reinforcement Learning: An Introduction}.
\newblock MIT press, 2018.

\bibitem[Gal and Ghahramani(2016)]{Gal:2016:Dropout}
Yarin Gal and Zoubin Ghahramani.
\newblock Dropout as a {Bayesian} approximation: {Representing} model
  uncertainty in deep learning.
\newblock In \emph{International Conference on Machine Learning}, pages
  1050--1059, 2016.

\bibitem[Coulom(2006)]{coulom2006efficient}
R{\'e}mi Coulom.
\newblock Efficient selectivity and backup operators in monte-carlo tree
  search.
\newblock In \emph{International Conference on Computers and Games}, pages
  72--83. Springer, 2006.

\bibitem[Browne et~al.(2012)Browne, Powley, Whitehouse, Lucas, Cowling,
  Rohlfshagen, Tavener, Perez, Samothrakis, and Colton]{browne2012survey}
Cameron~B Browne, Edward Powley, Daniel Whitehouse, Simon~M Lucas, Peter~I
  Cowling, Philipp Rohlfshagen, Stephen Tavener, Diego Perez, Spyridon
  Samothrakis, and Simon Colton.
\newblock A survey of {Monte Carlo} tree search methods.
\newblock \emph{IEEE Transactions on Computational Intelligence and AI in
  Games}, 4\penalty0 (1):\penalty0 1--43, 2012.

\bibitem[Silver et~al.(2017)Silver, Schrittwieser, Simonyan, Antonoglou, Huang,
  Guez, Hubert, Baker, Lai, Bolton, et~al.]{silver2017mastering}
David Silver, Julian Schrittwieser, Karen Simonyan, Ioannis Antonoglou, Aja
  Huang, Arthur Guez, Thomas Hubert, Lucas Baker, Matthew Lai, Adrian Bolton,
  et~al.
\newblock Mastering the game of go without human knowledge.
\newblock \emph{Nature}, 550\penalty0 (7676):\penalty0 354--359, 2017.

\bibitem[Pich{\'e} et~al.(2018)Pich{\'e}, Thomas, Ibrahim, Bengio, and
  Pal]{piche2018probabilistic}
Alexandre Pich{\'e}, Valentin Thomas, Cyril Ibrahim, Yoshua Bengio, and Chris
  Pal.
\newblock Probabilistic planning with sequential monte carlo methods.
\newblock In \emph{International Conference on Learning Representations}, 2018.

\bibitem[Tschantz et~al.(2020)Tschantz, Millidge, Seth, and
  Buckley]{tschantz2020control}
Alexander Tschantz, Beren Millidge, Anil~K Seth, and Christopher~L Buckley.
\newblock Control as hybrid inference.
\newblock \emph{arXiv preprint arXiv:2007.05838}, 2020.

\bibitem[Marino and Yue(2019)]{marinoinference}
Joseph Marino and Yisong Yue.
\newblock An inference perspective on model-based reinforcement learning.
\newblock \emph{ICML Workshop on Generative Modeling and Model-Based Reasoning
  for Robotics and AI}, 2019.

\bibitem[Van Der~Meer et~al.(2012)Van Der~Meer, Kurth-Nelson, and
  Redish]{van2012information}
Matthijs Van Der~Meer, Zeb Kurth-Nelson, and A~David Redish.
\newblock Information processing in decision-making systems.
\newblock \emph{The Neuroscientist}, 18\penalty0 (4):\penalty0 342--359, 2012.

\bibitem[Byers and Serences(2012)]{byers2012topdown}
Anna Byers and John~T. Serences.
\newblock Exploring the relationship between perceptual learning and top-down
  attentional control.
\newblock \emph{Vision Research}, 74:\penalty0 30 -- 39, 2012.

\bibitem[Higgins et~al.(2017)Higgins, Matthey, Pal, Burgess, Glorot, Botvinick,
  Mohamed, and Lerchner]{Higgins:2017:betaVAE}
Irina Higgins, Loic Matthey, Arka Pal, Christopher Burgess, Xavier Glorot,
  Matthew Botvinick, Shakir Mohamed, and Alexander Lerchner.
\newblock beta-vae: Learning basic visual concepts with a constrained
  variational framework.
\newblock \emph{International Conference on Learning Representations},
  2\penalty0 (5):\penalty0 6, 2017.

\bibitem[Matthey et~al.(2017)Matthey, Higgins, Hassabis, and
  Lerchner]{dsprites17}
Loic Matthey, Irina Higgins, Demis Hassabis, and Alexander Lerchner.
\newblock dsprites: Disentanglement testing sprites dataset.
\newblock https://github.com/deepmind/dsprites-dataset/, 2017.

\bibitem[Crosby et~al.(2019)Crosby, Beyret, and Halina]{Crosby:2019:animalai}
Matthew Crosby, Benjamin Beyret, and Marta Halina.
\newblock The {Animal-AI} olympics.
\newblock \emph{Nature Machine Intelligence}, 1\penalty0 (5):\penalty0
  257--257, 2019.

\bibitem[Kingma and Ba(2014)]{kingma2014adam}
Diederik~P Kingma and Jimmy Ba.
\newblock Adam: {A} method for stochastic optimization.
\newblock \emph{arXiv preprint arXiv:1412.6980}, 2014.

\bibitem[Hsieh et~al.(2018)Hsieh, Liu, Huang, Fei-Fei, and
  Niebles]{Hsieh:2018:nips}
Jun-Ting Hsieh, Bingbin Liu, De-An Huang, Li~F Fei-Fei, and Juan~Carlos
  Niebles.
\newblock Learning to decompose and disentangle representations for video
  prediction.
\newblock In \emph{Advances in Neural Information Processing Systems}, pages
  517--526, 2018.

\bibitem[Kim and Mnih(2018)]{Kim:2018:factorvae}
Hyunjik Kim and Andriy Mnih.
\newblock Disentangling by factorising.
\newblock \emph{arXiv preprint arXiv:1802.05983}, 2018.

\bibitem[Dhariwal et~al.(2017)Dhariwal, Hesse, Klimov, Nichol, Plappert,
  Radford, Schulman, Sidor, Wu, and Zhokhov]{baselines}
Prafulla Dhariwal, Christopher Hesse, Oleg Klimov, Alex Nichol, Matthias
  Plappert, Alec Radford, John Schulman, Szymon Sidor, Yuhuai Wu, and Peter
  Zhokhov.
\newblock Openai baselines.
\newblock \url{https://github.com/openai/baselines}, 2017.

\bibitem[Mnih et~al.(2015)Mnih, Kavukcuoglu, Silver, Rusu, Veness, Bellemare,
  Graves, Riedmiller, Fidjeland, Ostrovski, et~al.]{Mnih:2015:dqn}
Volodymyr Mnih, Koray Kavukcuoglu, David Silver, Andrei~A Rusu, Joel Veness,
  Marc~G Bellemare, Alex Graves, Martin Riedmiller, Andreas~K Fidjeland, Georg
  Ostrovski, et~al.
\newblock Human-level control through deep reinforcement learning.
\newblock \emph{nature}, 518\penalty0 (7540):\penalty0 529--533, 2015.

\bibitem[Mnih et~al.(2016)Mnih, Badia, Mirza, Graves, Lillicrap, Harley,
  Silver, and Kavukcuoglu]{Mnih:2016:A2C}
Volodymyr Mnih, Adria~Puigdomenech Badia, Mehdi Mirza, Alex Graves, Timothy
  Lillicrap, Tim Harley, David Silver, and Koray Kavukcuoglu.
\newblock Asynchronous methods for deep reinforcement learning.
\newblock In \emph{International conference on machine learning}, pages
  1928--1937, 2016.

\bibitem[Schulman et~al.(2017)Schulman, Wolski, Dhariwal, Radford, and
  Klimov]{Schulman:2017:PPO}
John Schulman, Filip Wolski, Prafulla Dhariwal, Alec Radford, and Oleg Klimov.
\newblock Proximal policy optimization algorithms.
\newblock \emph{arXiv preprint arXiv:1707.06347}, 2017.

\bibitem[Isomura and Friston(2018)]{isomura2018vitro}
Takuya Isomura and Karl~J Friston.
\newblock In vitro neural networks minimise variational free energy.
\newblock \emph{Nature Scientific Reports}, 8\penalty0 (1):\penalty0 1--14,
  2018.

\bibitem[Adams et~al.(2013)Adams, Shipp, and Friston]{adams2013predictions}
Rick~A Adams, Stewart Shipp, and Karl~J Friston.
\newblock Predictions not commands: Active inference in the motor system.
\newblock \emph{Brain Structure and Function}, 218\penalty0 (3):\penalty0
  611--643, 2013.

\bibitem[Kocsis and Szepesv{\'a}ri(2006)]{kocsis2006bandit}
Levente Kocsis and Csaba Szepesv{\'a}ri.
\newblock Bandit based {Monte-Carlo} planning.
\newblock In \emph{European Conference on Machine Learning}, pages 282--293.
  Springer, 2006.

\bibitem[Guo et~al.(2014)Guo, Singh, Lee, Lewis, and Wang]{guo2014deep}
Xiaoxiao Guo, Satinder Singh, Honglak Lee, Richard~L Lewis, and Xiaoshi Wang.
\newblock Deep learning for real-time {Atari} game play using offline
  {Monte-Carlo} tree search planning.
\newblock In \emph{Advances in Neural Information Processing Systems}, pages
  3338--3346, 2014.

\bibitem[Silver et~al.(2016)Silver, Huang, Maddison, Guez, Sifre, Van
  Den~Driessche, Schrittwieser, Antonoglou, Panneershelvam, Lanctot,
  et~al.]{silver2016mastering}
David Silver, Aja Huang, Chris~J Maddison, Arthur Guez, Laurent Sifre, George
  Van Den~Driessche, Julian Schrittwieser, Ioannis Antonoglou, Veda
  Panneershelvam, Marc Lanctot, et~al.
\newblock Mastering the game of go with deep neural networks and tree search.
\newblock \emph{Nature}, 529\penalty0 (7587):\penalty0 484, 2016.

\bibitem[{\c{C}}atal et~al.(2020){\c{C}}atal, Verbelen, Nauta, De~Boom, and
  Dhoedt]{ccatal2020learning}
Ozan {\c{C}}atal, Tim Verbelen, Johannes Nauta, Cedric De~Boom, and Bart
  Dhoedt.
\newblock Learning perception and planning with deep active inference.
\newblock In \emph{IEEE International Conference on Acoustics, Speech and
  Signal Processing}, pages 3952--3956, 2020.

\bibitem[Snider et~al.(2015)Snider, Lee, Poizner, and
  Gepshtein]{snider2015prospective}
Joseph Snider, Dongpyo Lee, Howard Poizner, and Sergei Gepshtein.
\newblock Prospective optimization with limited resources.
\newblock \emph{PLoS Computational Biology}, 11\penalty0 (9), 2015.

\bibitem[Solway and Botvinick(2015)]{solway2015evidence}
Alec Solway and Matthew~M Botvinick.
\newblock Evidence integration in model-based tree search.
\newblock \emph{Proceedings of the National Academy of Sciences}, 112\penalty0
  (37):\penalty0 11708--11713, 2015.

\bibitem[van Opheusden et~al.(2017)van Opheusden, Galbiati, Bnaya, Li, and
  Ma]{van2017computational}
Bas van Opheusden, Gianni Galbiati, Zahy Bnaya, Yunqi Li, and Wei~Ji Ma.
\newblock A computational model for decision tree search.
\newblock In \emph{CogSci.}, 2017.

\bibitem[Holding(1989)]{holding1989counting}
Dennis~H Holding.
\newblock Counting backward during chess move choice.
\newblock \emph{Bulletin of the Psychonomic Society}, 27\penalty0 (5):\penalty0
  421--424, 1989.

\bibitem[Huys et~al.(2012)Huys, Eshel, O'Nions, Sheridan, Dayan, and
  Roiser]{huys2012bonsai}
Quentin~JM Huys, Neir Eshel, Elizabeth O'Nions, Luke Sheridan, Peter Dayan, and
  Jonathan~P Roiser.
\newblock Bonsai trees in your head: {How} the {Pavlovian} system sculpts
  goal-directed choices by pruning decision trees.
\newblock \emph{PLoS Computational Biology}, 8\penalty0 (3), 2012.

\bibitem[Burns(2004)]{burns2004effects}
Bruce~D Burns.
\newblock The effects of speed on skilled chess performance.
\newblock \emph{Psychological Science}, 15\penalty0 (7):\penalty0 442--447,
  2004.

\bibitem[Van~Harreveld et~al.(2007)Van~Harreveld, Wagenmakers, and Van
  Der~Maas]{van2007effects}
Frenk Van~Harreveld, Eric-Jan Wagenmakers, and Han~LJ Van Der~Maas.
\newblock The effects of time pressure on chess skill: {An} investigation into
  fast and slow processes underlying expert performance.
\newblock \emph{Psychological Research}, 71\penalty0 (5):\penalty0 591--597,
  2007.

\bibitem[Mathieu et~al.(2018)Mathieu, Rainforth, Siddharth, and
  Teh]{Mathieu:2018}
Emile Mathieu, Tom Rainforth, N~Siddharth, and Yee~Whye Teh.
\newblock Disentangling disentanglement in variational autoencoders.
\newblock \emph{arXiv preprint arXiv:1812.02833}, 2018.

\bibitem[Kim et~al.(2019)Kim, Wang, Sahu, and Pavlovic]{Kim:2019}
Minyoung Kim, Yuting Wang, Pritish Sahu, and Vladimir Pavlovic.
\newblock {Bayes-Factor-VAE: Hierarchical Bayesian} deep auto-encoder models
  for factor disentanglement.
\newblock In \emph{IEEE International Conference on Computer Vision}, pages
  2979--2987, 2019.

\bibitem[Fatemi~Shariatpanahi and Nili~Ahmadabadi(2007)]{hadi2007attention}
Hadi Fatemi~Shariatpanahi and Majid Nili~Ahmadabadi.
\newblock Biologically inspired framework for learning and abstract
  representation of attention control.
\newblock In \emph{Attention in Cognitive Systems. Theories and Systems from an
  Interdisciplinary Viewpoint}, pages 307--324, 2007.

\bibitem[Mott et~al.(2019)Mott, Zoran, Chrzanowski, Wierstra, and
  Rezende]{mott2019towards}
Alexander Mott, Daniel Zoran, Mike Chrzanowski, Daan Wierstra, and Danilo~J
  Rezende.
\newblock Towards interpretable reinforcement learning using attention
  augmented agents.
\newblock In \emph{Advances in Neural Information Processing Systems}, pages
  12329--12338, 2019.

\bibitem[Parr et~al.(2018)Parr, Benrimoh, Vincent, and
  Friston]{parr2018perceptual}
Thomas Parr, David~A. Benrimoh, Peter Vincent, and Karl~J Friston.
\newblock Precision and false perceptual inference.
\newblock \emph{Frontiers in Integrative Neuroscience}, 12:\penalty0 39, 2018.

\bibitem[Dayan et~al.(2000)Dayan, Kakade, and Montague]{dayan2000learning}
Peter Dayan, Sham Kakade, and Read~P Montague.
\newblock Learning and selective attention.
\newblock \emph{Nature Neuroscience}, 3\penalty0 (11):\penalty0 1218--1223,
  2000.

\bibitem[Baluch and Itti(2011)]{baluch2011mechanisms}
Farhan Baluch and Laurent Itti.
\newblock Mechanisms of top-down attention.
\newblock \emph{Trends in Neurosciences}, 34\penalty0 (4):\penalty0 210--224,
  2011.

\bibitem[Sasaki et~al.(2010)Sasaki, Nanez, and Watanabe]{sasaki2010advances}
Yuka Sasaki, Jose~E Nanez, and Takeo Watanabe.
\newblock Advances in visual perceptual learning and plasticity.
\newblock \emph{Nature Reviews Neuroscience}, 11\penalty0 (1):\penalty0 53--60,
  2010.

\bibitem[Posner and Petersen(1990)]{posner1990attention}
Michael~I Posner and Steven~E Petersen.
\newblock The attention system of the human brain.
\newblock \emph{Annual Review of Neuroscience}, 13\penalty0 (1):\penalty0
  25--42, 1990.

\bibitem[Dayan and Yu(2002)]{dayan2001nips}
Peter Dayan and Angela~J Yu.
\newblock {ACh}, uncertainty, and cortical inference.
\newblock In \emph{Advances in Neural Information Processing Systems}, pages
  189--196, 2002.

\bibitem[Gu(2002)]{gu2002neuromodulatory}
Q~Gu.
\newblock Neuromodulatory transmitter systems in the cortex and their role in
  cortical plasticity.
\newblock \emph{Neuroscience}, 111\penalty0 (4):\penalty0 815--835, 2002.

\bibitem[Yu and Dayan(2005)]{Yu:2005:Uncertainty}
Angela~J Yu and Peter Dayan.
\newblock Uncertainty, neuromodulation, and attention.
\newblock \emph{Neuron}, 46\penalty0 (4):\penalty0 681--692, 2005.

\bibitem[Moran et~al.(2013)Moran, Campo, Symmonds, Stephan, Dolan, and
  Friston]{moran2013free}
Rosalyn~J Moran, Pablo Campo, Mkael Symmonds, Klaas~E Stephan, Raymond~J Dolan,
  and Karl~J Friston.
\newblock Free energy, precision and learning: {The} role of cholinergic
  neuromodulation.
\newblock \emph{Journal of Neuroscience}, 33\penalty0 (19):\penalty0
  8227--8236, 2013.

\bibitem[Parr(2019)]{parr2019computational}
Thomas Parr.
\newblock \emph{The Computational Neurology of Active Vision}.
\newblock PhD thesis, University College London, 2019.

\bibitem[Fellows et~al.(2019)Fellows, Mahajan, Rudner, and
  Whiteson]{fellows2019virel}
Matthew Fellows, Anuj Mahajan, Tim~GJ Rudner, and Shimon Whiteson.
\newblock Virel: A variational inference framework for reinforcement learning.
\newblock In \emph{Advances in Neural Information Processing Systems}, pages
  7122--7136, 2019.

\bibitem[Levine(2018)]{levine2018reinforcement}
Sergey Levine.
\newblock Reinforcement learning and control as probabilistic inference:
  Tutorial and review.
\newblock \emph{arXiv preprint arXiv:1805.00909}, 2018.

\bibitem[Parr and Friston(2018)]{parr2018computational}
Thomas Parr and Karl~J Friston.
\newblock The computational anatomy of visual neglect.
\newblock \emph{Cerebral Cortex}, 28\penalty0 (2):\penalty0 777--790, 2018.

\end{thebibliography}

\newpage

\section{Supplementary material}

\subsection{Expected free energy derivation}

Here, we provide the steps needed to derive Eq.~\eqref{eq:G5} from Eq.~\eqref{eq:G3}. The term \eqref{eq:G3b} can be re-written as:
\begin{align*}
\mathop{\mathbb{E}_{\tilde{Q}}}\big[ &\log Q(s_{\tau}|\pi) - \log Q(s_{\tau}|o_{\tau},\pi) \big]=\\
&\;\;\;\;=\mathop{\mathbb{E}_{Q(\theta|\pi)Q(s_{\tau}|\theta,\pi)Q(o_{\tau}|s_{\tau},\theta, \pi)}}\big[ \log Q(s_{\tau}|\pi) - \log Q(s_{\tau}|o_{\tau},\pi) \big]\\
&\;\;\;\;=\mathop{\mathbb{E}_{Q(\theta|\pi)}}\Big[\mathop{\mathbb{E}_{Q(s_{\tau}|\theta,\pi)}} \log Q(s_{\tau}|\pi) - \mathop{\mathbb{E}_{Q(s_{\tau}, o_{\tau}|\theta,\pi)}} \log Q(s_{\tau}|o_{\tau},\pi) \Big]\\
&\;\;\;\;=\mathop{\mathbb{E}_{Q(\theta|\pi)}}\Big[\E{Q(o_\tau | \theta, \pi)} H(s_{\tau}|o_{\tau},\pi) - H(s_{\tau}|\pi) \Big] ~ ,
\end{align*}
where we have only used the definition $\tilde{Q} = Q(o_\tau, s_\tau, \theta | \pi)$, and the definition of the standard (and conditional) Shannon entropy. 

Next, the term \eqref{eq:G3c} can be re-written as:
\begin{align*}
\mathop{\mathbb{E}_{\tilde{Q}}}\big[ \log Q(\theta|s_{\tau},\pi) &- \log Q(\theta|s_{\tau},o_{\tau},\pi) \big]=\\
&\;\;\;\;=\mathop{\mathbb{E}_{Q(s_{\tau},\theta, o_{\tau}|\pi)}}\big[ \log Q(o_{\tau}|s_{\tau},\pi) - \log Q(o_{\tau}|s_{\tau},\theta,\pi) \big]\\
&\;\;\;\;=\mathop{\mathbb{E}_{Q(s_{\tau}|\pi)Q(o_{\tau}|s_{\tau},\pi)}} \log Q(o_{\tau}|s_{\tau},\pi)\\
&\;\;\;\;\;\;\;\;\;\;\;\; - \mathop{\mathbb{E}_{Q(\theta|\pi)Q(s_{\tau}|\theta,\pi)Q(o_{\tau}|s_{\tau},\theta\pi)}}\log Q(o_{\tau}|s_{\tau},\theta,\pi)\\
&\;\;\;\;= \mathop{\mathbb{E}_{Q(\theta|\pi)Q(s_{\tau}|\theta,\pi)}}H(o_{\tau}|s_{\tau},\theta,\pi) - \mathop{\mathbb{E}_{Q(s_{\tau}|\pi)}} H(o_{\tau}|s_{\tau},\pi) ~ ,
\end{align*}
where the first equality is obtained via a normal Bayes inversion, and the second via the factorization of $\tilde{Q}$. These two terms can be directly combined to obtain Eq.~\eqref{eq:G5}.
With this expression at hand, the only problem that remains is estimating these quantities from the outputs of all neural networks involved. We provide some information here, in addition to that in Sec.~\ref{sec:calculating}.

For Eq.~\eqref{eq:G5b}, $H(s_\tau\!\mid\!\pi)$ is estimated sampling from the transition network, and $H(s_\tau\!\mid\!o_\tau, \pi)$ from the encoder network (both parameterised with Gaussians, so entropies can be calculated from log-variances).
For the first term in Eq.~\eqref{eq:G5c} we sample several $\theta$ from the MC-dropouts and several $s_\tau$ from the transition network; then average the entropies $H(o_\tau\!\mid\!s_\tau,\theta,\pi)$ (which are closed-form since $o_\tau$ is Bernoulli-distributed) over the $(\theta,s_\tau)$ samples. For the second term, we fix the $\theta$ and sample multiple $s_\tau$ (so that, effectively, $p(o|s) = \sum_\theta p(o|s,\theta) p(\theta)$ is approximated with a single MC sample) and repeat the procedure.
Although noisy, this estimator was found to be fast and suitable for training.
Finally, note that in both cases the quantities computed correspond to a difference between the entropy of the average and the average of the entropies -- which is the mutual information%
, a known part of the EFE.

\newpage

\subsection{Glossary of terms and notation}

\begin{table}[htbp!]
\begin{tabular}{|l|p{0.5\linewidth}|}
\hline
\multicolumn{1}{|c|}{\textbf{Notation}}& \multicolumn{1}{c|}{\textbf{Definition}}                      
\tabularnewline \hline
\hline
 $S$ & Set of all possible hidden states  \tabularnewline 
 $s_t$ & Hidden state at time $t$, random variable over $S$  \tabularnewline 
 $s_{1:t}$ & Sequence of hidden states, $s_1,.., s_t$, random variable over $S^t$  \tabularnewline 
 $O$ & Set of all possible observations  \tabularnewline
 $o_t$ & Observation at time $t$, random variable over $O$  \tabularnewline 
 $o_{1:t}$ & Sequence of observations, $o_1,.., o_t$, random variable over $O^t$  \tabularnewline 
 $T$ & Number of time steps in episode, positive integer  \tabularnewline
 $U$ & Set of all possible actions  \tabularnewline 
 $a_t$ & Action at time $t$, random variable over $U$  \tabularnewline 
 $\Pi$ & Set of all allowable policies; i.e., sequences of actions, subset of $U^t$   \tabularnewline 
 $\pi$ & Policy as defined by $(a_1,a_2,...,a_T)$, random variable over $\Pi$ \tabularnewline
 $P_{\theta_s}(s_t|s_{t-1},a_{t-1})$ & Transition function; parameterized by  $\theta_s$ \tabularnewline 
 $P_{\theta_o}(o_t|s_{t})$ & Likelihood/observation function; parameterized by $\theta_o$ \tabularnewline 
 $P_{\theta_{o,s}}(s_{1:T}, a_{1:T-1},o_{1:T})$ & Generative model; factorized form $P(a_1)P(s_1)$ $\prod^T_{t=2}P_{\theta_s}(s_t|s_{t-1},a_{t-1})\prod^T_{t=1}P_{\theta_o}(o_t|s_t)$ \tabularnewline
 $Q_{\phi_a}(a_t)$ & Approximate posterior over actions; parameterized by $\phi_a$. Dependency on $s_t$ has been dropped following standard variational inference notation.\tabularnewline
 $Q_{\phi_s}(s_t)$ & Approximate posterior over hidden states; parameterized by $\phi_s$. Dependency on $o_t$ has been dropped following standard variational inference notation.\tabularnewline
 $Q_{\phi}(s_{1:T}, a_{1:T-1})$ & Approximate posterior over actions and hidden states with mean-field assumptions; $\prod^T_{t=1}Q_{\phi_s}(s_t)Q_{\phi_a}(a_t)$ \tabularnewline
 $-\log P_{\theta}(o_{t})$ &  Negative log-likelihood; surprisal at time $t$ \tabularnewline
 $\mathop{\mathbb{E}_{Q_{\phi}(s_{t}, a_{t})}}\big[ \log Q_{\phi}(s_{t}, a_{t}) - \log P_{\theta}(o_{t}, s_{t}, a_{t}) \big]$ & Variational free energy or evidence lower bound at time $t$. This can be decomposed to Eq.~\eqref{eq:Pa} using the appropriate factorization. \tabularnewline
 $Q_{\phi}(s_{1:T}, a_{1:T-1},o_{1:T})$ & Approximate posterior with mean-field assumptions; $\prod^T_{t=1}Q_{\phi_s}(s_t)Q_{\phi_a}(a_t)P_{\theta_o}(o_t|s_{t})$ \tabularnewline
 $\mathop{\mathbb{E}_{P(o_{\tau}|s_{\tau},\theta)}\mathbb{E}_{Q_{\phi}(s_{\tau},\theta|\pi)}}\big[ \log Q_{\phi}(s_{\tau},\theta|\pi)$ & Expected free energy, defined on $\Pi$, for some \tabularnewline
 $- \log P(o_{\tau},s_{\tau},\theta|\pi) \big]$
 & future time--point $\tau$. This is derived by taking an additional expectation $P(o_\tau|s_\tau, \theta)$ where $\theta$ denotes random variable over learnt distribution $P_\theta(.|\pi)$ \tabularnewline
 $\sigma$ & Softmax function or normalized exponential \tabularnewline
 $P(\pi)$ &  Posterior distribution about policies via softmax function of the summed (negative) expected free energy over time; $\sigma\Big(- \sum_{\tau>t}{G(\pi,\tau)}\Big)$ \tabularnewline
 $P(a_t)$ & Posterior distribution about actions via summed probability of all policies that begin with a particular action, $a_t$ \tabularnewline
  \hline
\end{tabular}
\caption{Glossary of terms and notation}\label{gloss}
\label{table}   
\end{table}

\newpage

\subsection{Training Procedure}

The model presented here was implemented in Python and the library TensorFlow 2.0. We initialized 3 different ADAM optimizers, which we used in parallel, to allow learning parameters with different rates. The networks $Q_{\phi_s},P_{\theta_o}$ were optimized using an initial learning rate of $10^{-3}$ and, as a loss function, the first two terms of Eq.~\eqref{eq:Ft}. In experiments where regularization was used, the loss function used by this optimizer was adjusted to
\begin{align}
\begin{split}\label{eq:F:regularization}
L_{\phi_s,\theta_o} = &-\E{Q(s_t)}\big[ \log P(o_t|s_t; \theta_o) \big] + \gamma \KL{Q_{\phi_s}(s_t)}{P(s_t|s_{t-1},a_{t-1}; \theta_s)} \\
& + (1-\gamma) \KL{Q_{\phi_s}(s_t)}{N(0,1)} ~ ,
\end{split}
\end{align}
where $\gamma$ is a hyper parameter, starting with value $0$ and gradually increasing to $0.8$. In our experiments, we found that the effect of regularization is only to improve the speed of convergence and not the behavior of the agent and, thus, it can be safely omitted.

The parameters of the network $P_{\theta_s}$ were optimized using a rate of $10^{-4}$ and only the second term of Eq.~\eqref{eq:Ft} as a loss. Finally, the parameters of $Q_{\phi_a}$ were optimized with a learning rate of $10^{-4}$ and only the final term of Eq.~\eqref{eq:Ft} as a loss. For all presented experiments and learning curves, batch size was set to 50. A learning iteration is defined as 1000 optimization steps with new data generated from the corresponding environment.

In order to learn to plan further into the future, the agents were trained to map transitions every 5 simulation time-steps in dynamic dSprites and 3 simulation time-steps in Animal-AI. Finally, the runtime of the results presented here is as follows. For the agents in the dynamic dSprites environment, training of the final version of the agents took approximately $26$ hours per version (on-policy, 700 learning iterations) using an NVIDIA Titan RTX GPU. Producing the learning and performance curves in Fig.~\ref{fig:results:dsprites}, took $10$ hours per agent when the 1-step and habitual strategies were employed and approximately $4$ days when the full MCTS planner was used (Fig.~\ref{fig:results:dsprites}A).
For the Animal-AI environment, off-policy training took approximately $9$ hours per agent, on-policy training took $8$ days and, the results presented in Fig.~\ref{fig:planning} took approximately $4$ days, using an NVIDIA GeForce GTX 1660 super GPU (CPU: i7-4790k, RAM: 16GB DDR3).

\subsection{Training algorithm}

The following algorithm is described for a single environment ($batch=1$), to maintain notation consistency with the main text, but can also be applied when $batch>1$. This algorithm is exactly the same for both Dynamic dSprites and Animal-AI environments. Finally, for either off-policy or off-line training, the action applied to the environment (line 9) is drawn from a different policy or loaded from a pre-recorded data-set respectively.

\begin{algorithm}
	\caption{DAIMC on-policy training}
	\begin{algorithmic}[1]
    		\For {$t=1,2,\ldots,\text{max iterations}$}
    		\State Randomize environment and sample a new observation $\tilde{o}_t$. 
            \State Run planner and compute prior policy $P(a_t)$.
		    \State Compute $Q_{\phi_s}(s_t)$ using $\tilde{o}_t$.
		    \State Compute $Q_{\phi_a}(a_t)$ using a sampled state $\tilde{s}_t \sim Q_{\phi_s}(s_t)$.
		    \State Compute $D_t = \KL{Q_{\phi_a}(a_t)}{P(a_t)}$.
		    \State Apply a gradient step on $\phi_a$ using $D_t$ as loss.
		    \State Compute $\omega_{t+1}$ from Eq.~\eqref{eq:precision} using $D_t$.
	        \State Apply action $\tilde{a_t} \sim P(a_t)$ to the environment and sample a new observation $\tilde{o}_{t+1}$.
		    \State Compute $\mu,\sigma$ from $P_{\theta_s}(s_{t+1}|\tilde{s_t},\tilde{a_t})$.
		    \State Compute $Q_{\phi_s}(s_{t+1})$ using $\tilde{o}_{t+1}$.
		    \State Apply a gradient step on $\theta_s$ using $\KL{Q_{\phi_s}(s_{t+1})}{
		    \mathcal{N}(\mu, \sigma^2/{\omega_t})}$.
		    \State Apply a gradient step on $\phi_s$, $\theta_o$ using $-\E{Q(s_{t+1})}\big[ \log P_{\theta_o}(o_{t+1}|s_{t+1}) \big] + \KL{Q_{\phi_s}(s_{t+1})}{
		    \mathcal{N}(\tilde{\mu}, \tilde{\sigma}^2/{\omega_t})}$.
		\EndFor
	\end{algorithmic} 
\end{algorithm}

\newpage

\subsection{Model parameters}

\begin{figure}[h!]
\includegraphics[width=0.8\linewidth]{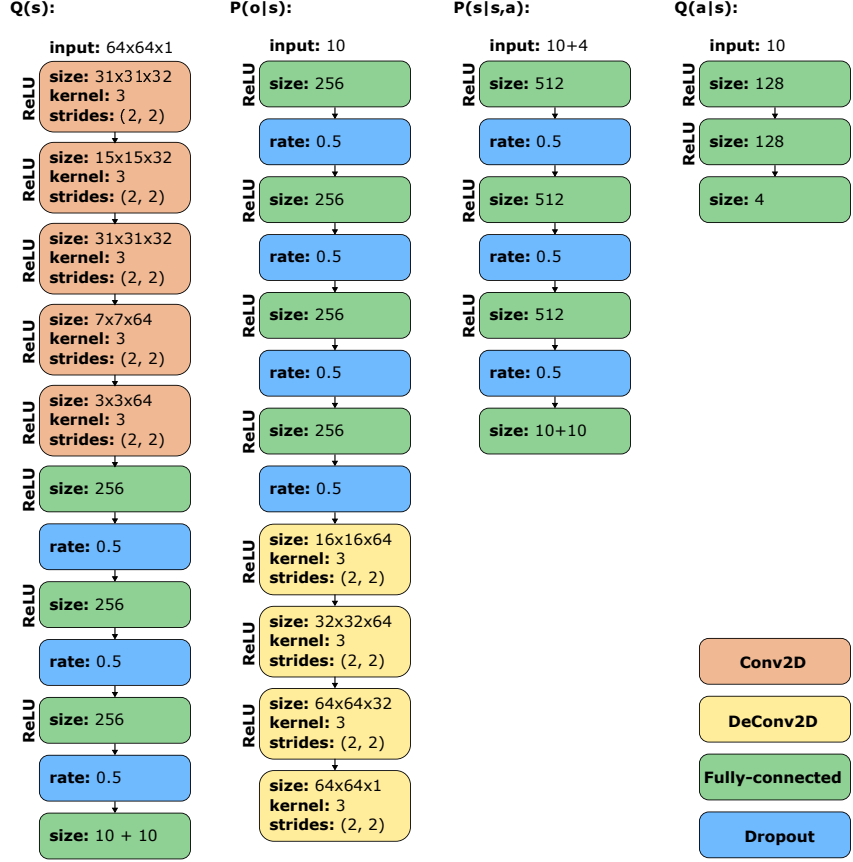}
\centering
\caption{Neural network parameters used for the dynamic dSprites experiments. For the Animal-AI experiments, the only differences are: i) the input layer of the network used for $Q_{\phi_s}(s)$ and output layer for $P_{\theta_o}(o_t|s_t)$ have shape $(32,32,3)$, ii) the input layer of $P_{\theta_s}(s_{t+1}|s_t,a_t)$ has shape $(10+3)$ and iii) the output layer of $Q_{\phi_a}(a_t)$ has a shape of $(3)$, corresponding to the three actions forward, left and right.}
\label{fig:net_parameters}
\end{figure}

In both simulated environments, the network structure used was almost identical, consisting of convolutional, deconvolutional, fully-connected and dropout layers (Fig.~\ref{fig:net_parameters}). In both cases, the dimensionality of the latent space $s$ was 10. For the top-down attention mechanism, the parameters used were $\alpha = 2, b = 0.5, c = 0.1$ and $d = 5$ for the Animal-AI environment and $\alpha = 1, b = 25, c = 5$ and $d = 1.5$ for dynamic dSprites. The action space was $|\mathcal{A}|=3$ for Animal-AI and $|\mathcal{A}|=4$ for dynamic dSprites.
Finally, with respect to the planner, we set $c_{\text{explore}}=1$ in both cases, $T_{dec}=0.8$ (when another value is not specifically mentioned), the depth of MCTS simulation rollouts was set to $3$, while the maximum number of MCTS loops was set to $300$ for dynamic dSprites and $100$ for Animal-AI.

\newpage

\subsection{Examples of agent plans}

\begin{figure}[h!]
\includegraphics[width=0.98\linewidth]{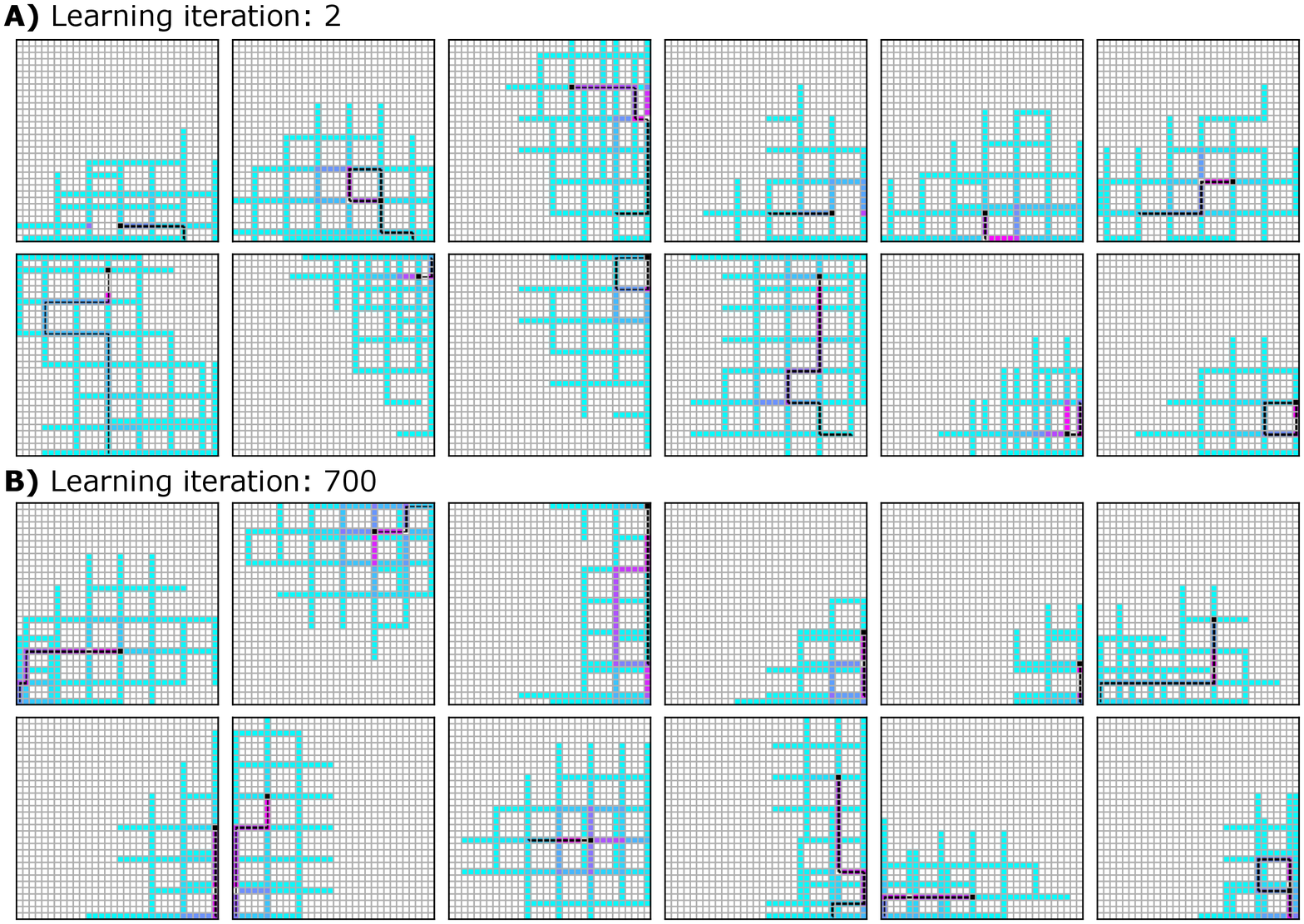}
\centering
\caption{Examples of consecutive plans in the dynamic dSprites environment during a single experiment. }
\label{fig:examples:dsprites}
\end{figure}

\begin{figure}[h!]
\includegraphics[width=0.98\linewidth]{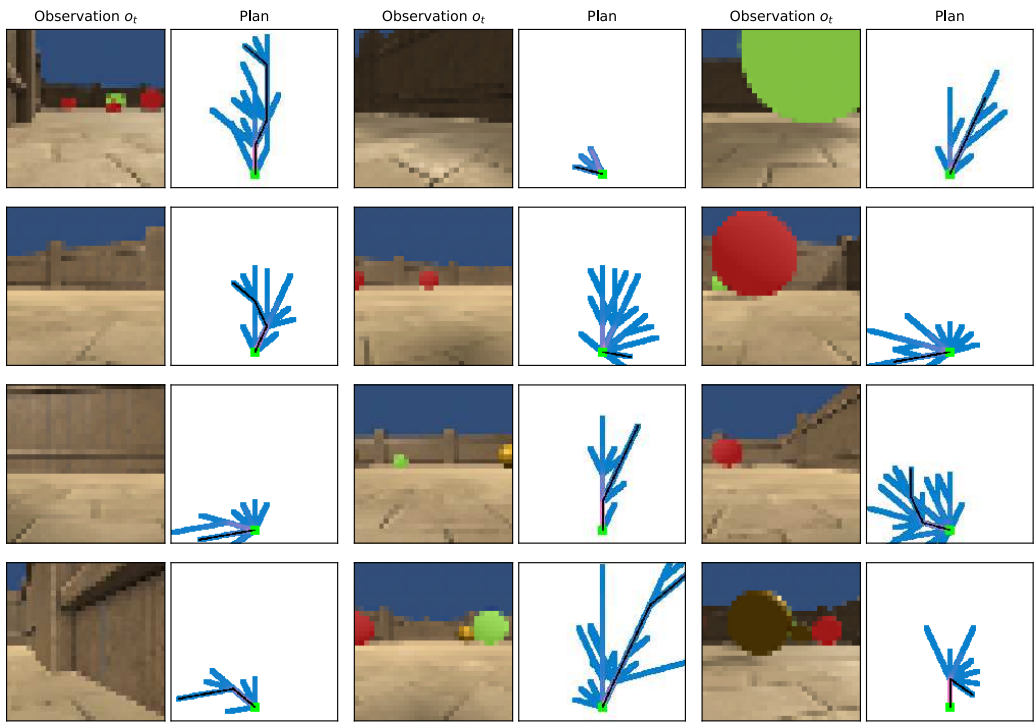}
\centering
\caption{Examples of plans in Animal-AI environment. Examples were picked randomly. }
\label{fig:examples:animalai}
\end{figure}

\newpage

\subsection{Examples of traversals}

\begin{figure}[h!]
\includegraphics[width=\linewidth]{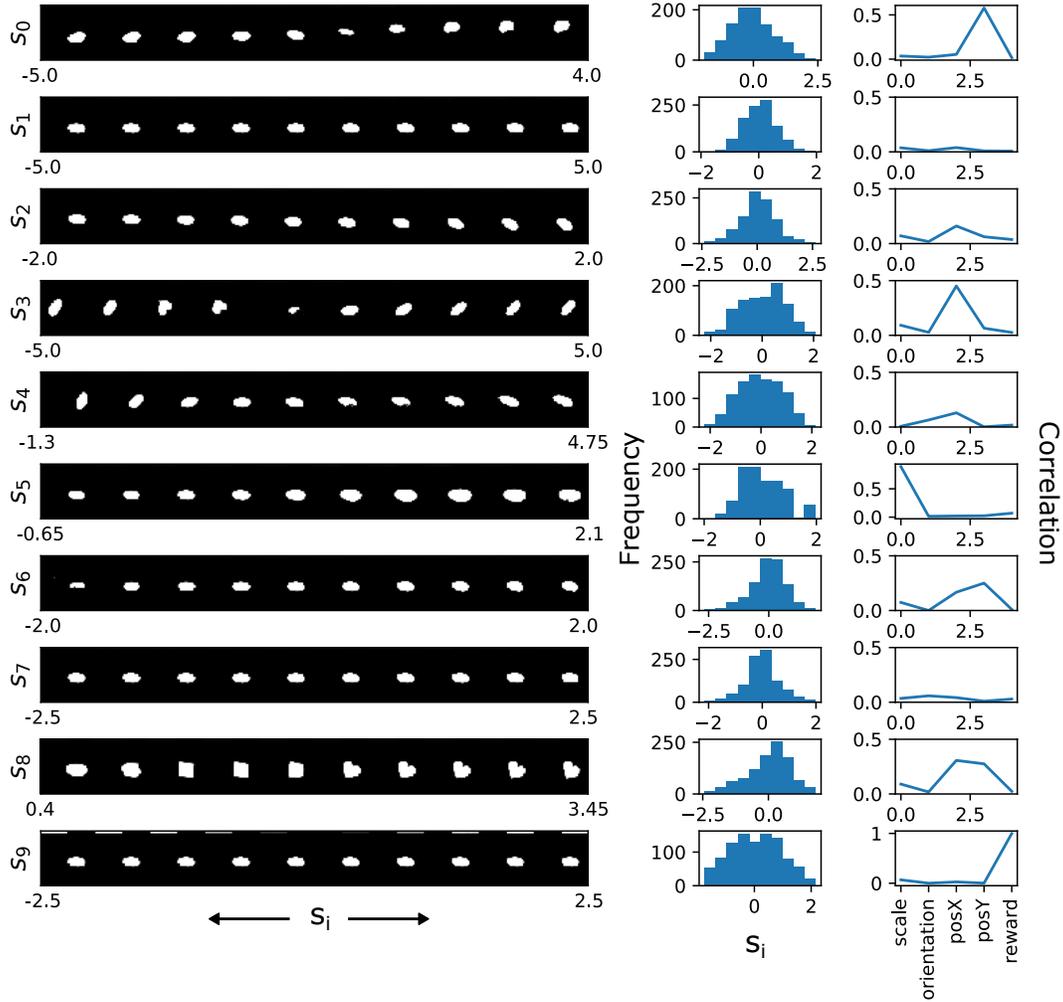}
\centering
\caption{Latent space traversals for the full active inference agent optimized in the dynamic dSprites environment. Histograms represent distribution of values for 1000 random observations. The graphs on the right column represent correlation between each dimension of $s$ and the 6 ground truth features of the environment. This includes the 5 features of the dSprites dataset and reward, encoded in the top pixels shown in $s_9$. }
\label{fig:examples:traversals}
\end{figure}

\end{document}